\documentclass[twocolumn,showpacs,aps,superscriptaddress,floatfix,aps]{revtex4}
\usepackage{amsmath,amssymb,eucal,hyperref,graphicx}
\begin{document}
\title{Escape of a Uniform Random Walk from an Interval}

\author{T.~Antal}\affiliation{Center of Polymer Studies
  and Department of Physics, Boston University, Boston, Massachusetts, 02215
  USA} \author{S.~Redner}\email{redner@bu.edu} \affiliation{Center of Polymer
  Studies and Department of Physics, Boston University, Boston,
  Massachusetts, 02215 USA}
\begin{abstract}
  
  We study the first-passage properties of a random walk in the unit interval
  in which the length of a single step is uniformly distributed over the
  finite range $[-a,a]$.  For $a$ of the order of one, the exit probabilities
  to each edge of the interval and the exit time from the interval exhibit
  anomalous properties stemming from the change in the minimum number of
  steps to escape the interval as a function of the starting point.  As $a$
  decreases, first-passage properties approach those of continuum diffusion,
  but non-diffusive effects remain because of residual discreteness effects.

\end{abstract}
\pacs{02.50.C2, 05.40.Fb}

\maketitle

\section{Introduction}

Consider a discrete-time random walk in which the length of each step is
uniformly distributed in the range $[-a,a]$.  We term this process the
uniform random walk (URW).  The walker is initially at an arbitrary point $x$
in the unit interval [0,1] and the endpoints are absorbing.  For the URW, an
absorbing boundary is defined such that if the walk crosses an endpoint of
the interval, the walk is trapped exactly at this endpoint.  We are
interested in the first-passage properties of this URW.

One motivation for this study comes from the problem of DNA sequence
recognition by a mobile protein \cite{general,C}.  The protein molecule
typically seeks its target by a combination of diffusion along the DNA chain,
and also detachment and subsequent reattachment of the protein at a point
many base pairs away, and the basic quantity of interest is the time required
for the protein to find its target on a finite DNA chain \cite{general,C,M}.
The two mechanisms of diffusion and detachment/reattachment can be viewed as
a random walk along the chain with a variable step length distribution
\cite{M}.  This is the viewpoint that we shall adopt for this work.  A second
motivation for our work is that when individual step lengths are drawn from a
continuous distribution, the resulting random walk exhibit a variety of
interesting properties beyond those of the discrete random walk.  These
include, among others, both unusual first-passage properties \cite{MM,DEH} as
well as extreme-value phenomena \cite{MCZ}.

Here we study the related problem of first-passage properties of the URW in a
finite interval.  Perhaps the most basic such feature is the exit probability
$R(x)$, defined as the probability for a walk that starts at $x$ to
eventually cross the boundary at $x=1$.  Since the probability to exit via
the left boundary is $1-R(x)$, we need only consider exit to the right
boundary, for which we reserve the term exit probability.  A related quantity
is the mean exit time $t(x)$, defined as the average time to exit the
interval at either boundary when the walk starts from $x$.  This exit time
from an interval has been investigated for a random walk with general
single-step hopping probabilities \cite{LS,L,WS,BPBW} and also in an
econophysics context \cite{MM}.  As we shall see, the interplay between the
maximum step size and the interval length leads to commensuration effects
that are absent when the step length distribution extends over an infinite
range.

For both pure diffusion and the classical random walk with the step length
$\Delta x=1/N$, where $N$ is an arbitrary integer, it is well known that the
exit probability from an absorbing interval is $R(x)=x$ \cite{feller,fpp}.
Similarly, the mean exit time is $t(x)=x(1-x)/2D$, where the diffusion
coefficient for the discrete random walk is $D=\langle (\Delta
x)^2\rangle/2$, with $\langle (\Delta x)^2\rangle$ the mean-square length of
a single step.  We now determine these first-passage properties---the exit
probability and the exit time---for the URW.

\section{General Features}
\label{general}

To help visualize the general behavior, we performed numerical simulations of
the URW by a probability propagation algorithm; for the URW, this approach is
orders of magnitude more efficient than direct Monte Carlo simulation. In
probability propagation, we first divide the unit interval into $N$ discrete
points.  We correspondingly discretize the URW as follows: a URW that moves
uniformly within the range $[-a,a]$ at each step is equivalent to its
discretized counterpart hopping equiprobably to any one of the $aN$ discrete
points on either side of the current site.  When an element of probability
hops outside the interval, this element is considered to be trapped at the
boundary where the element left the interval.  We continue this propagation
until less than $10^{-6}$ of the initial probability remains in the interval.
Running the propagation further led to insignificant corrections.  The
simulations were generally performed with $N=5000$.  We found negligible
differences in our results when the interval was discretized into $N=10000$
points.  This probability propagation algorithm is also considerably more
efficient than naive Monte Carlo simulation of an ensemble of random walkers.

\begin{figure}[ht]
 \vspace*{0.cm}
\includegraphics*[width=0.4\textwidth]{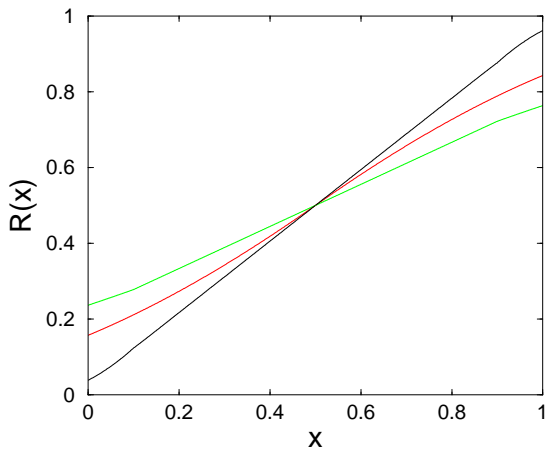}\vskip 0.1in 
\caption{Probability of exit at the right boundary, $R(x)$, versus 
  $x$ for $a=0.9$, 0.5, and 0.1 (slopes increasing, respectively), where $a$
  is the width of the single-step distribution.  The data for $a=0.1$ are
  based on probability propagation, while the other two datasets are obtained
  analytically.
  \label{exit}}
\end{figure}

\begin{figure}[ht]
 \vspace*{0.cm}
\includegraphics*[width=0.4\textwidth]{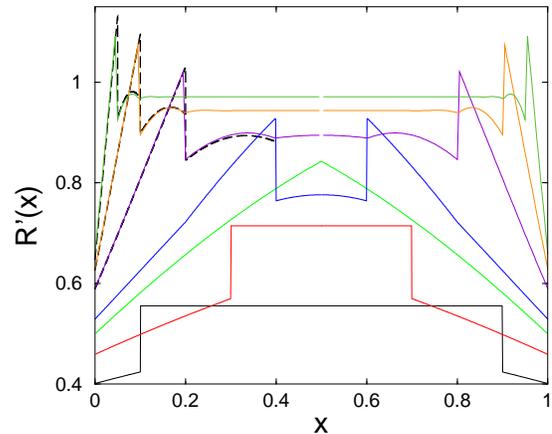}\vskip 0.1in 
\caption{The derivative $R'(x)$ versus $x$ for $a=0.9$, 0.7, 0.5, 0.4, 0.2, 
  0.1, and 0.05 (rising curves, respectively).  Shown dashed are the results of
    an asymptotic approximation that is discussed in Sec.~\ref{small-a}.
  \label{rprime}}
\end{figure}

When $a$ is of the order of 1, non-diffusive features arise because the walk
can traverse the interval in just a few steps.  In the limit $a\to\infty$,
$R(x)=1/2$ for all $x$; that is, either boundary is reached equiprobably,
independent of the initial position.  Conversely, for $a\ll 1$, the
first-passage properties of the URW approach those of continuum diffusion.
Thus $R(x)\approx x$, except when the starting point is close to a boundary
(Fig.~\ref{exit}).  In this boundary region, $R(x)$ undergoes a series of
transitions in which the $n^{\rm th}$ derivative is discontinuous whenever
$x$ or $1-x$ passes through $na$.  These transitions become more apparent
upon plotting $R'(x)$ versus $x$ (Fig.~\ref{rprime}).  For small $a$, the
qualitative behavior of $R'$ is reminiscent of the Gibbs' overshoot
phenomenon \cite{Arf} when expanding a square wave in a Fourier series.

It is worth noting that the URW in an absorbing interval is not a martingale
\cite{martin} and hence $R(x)\ne x$ in general.  That is, the URW is not a
``fair'' process.  The unfairness arises because when a walk crosses the
boundary and ostensibly lands outside the interval, the walk is reassigned to
be trapped at the edge of the interval.  Thus the mean position of the
probability distribution is not conserved.  Another consequence of the
unfairness is that the exit probability $R(x=0)$ for fixed $a$, no matter how
small, is non-zero.  The possibility of starting at $x=0$ and exiting at
$x=1$ can be viewed as an effective bias of the walk toward the middle of
the interval.  In the context of the gambler's ruin problem \cite{feller}, a
gambler that is about to be ruined appears to be best served by making a
reckless bet that is of the order of the total amount of capital in the game.

\begin{figure}[ht]
 \vspace*{0.cm}
\includegraphics*[width=0.4\textwidth]{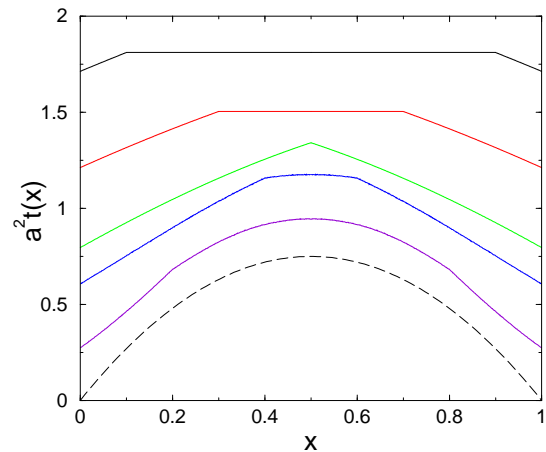}\vskip 0.0in 
\caption{Mean time to exit at either boundary, $t(x)$, times $a^2$
  (proportional to the diffusion coefficient) versus $x$ for $a\to 0$,
  $a=0.2$, 0.4, 0.5, 0.7 and 0.9 (bottom to top).  The data for $a<0.5$ is
  based on simulation, while the remaining data is obtained analytically.
  \label{exit-time}}
\end{figure}

The mean exit time of the URW deviates strongly from the continuum diffusive
form $t(x)=x(1-x)/2D$, when $a$ becomes of the order of 1
(Fig.~\ref{exit-time}).  Notice also that $t(x)$ does not go to zero as $x\to
0$ or $x\to 1$.  This limiting behavior again reflects the fact that there
is a non-negligible probability for a walk that starts at one edge of the
interval to exit via the opposite edge, a process that requires a non-zero
time.

\section{First-Passage Properties}

In general, the exit probability may be determined from the backward equation
that expresses the exit probability $R(x)$ in terms of the exit probability
after one step of the random walk has elapsed \cite{fpp}.  This backward
equation has the generic form
\begin{equation}
\label{back-r}
R(x)=\int dx'\, p(x\to x')R(x')\,.
\end{equation}
That is, the exit probability starting from $x$ equals the probability of
making a single step to $x'$ times the probability of exit from $x'$,
integrated over all possible values of $x'$.  In a parallel fashion, the mean
exit time can generically be written as
\begin{equation}
\label{back-t}
t(x)=\int dx'\,  p(x\to x')[t(x')+1]\,.
\end{equation}
That is, the exit time starting from $x$ equals one plus the exit time from
$x'$, when integrated over all possible values of $x'$, with each term
weighted by $p(x\to x')$.  Note that the trailing factor of 1 can be taken
outside the integral sum since $\int dx'\, p(x\to x')=1$.  We now apply these
two formulae to determine first-passage properties for the URW by studying,
in turn, the cases $a>1$, $a\in [1/2,1]$, $a\in [1/3,1/2]$, $a<1/3$, and
finally $a\to 0$.

\subsection{$a>1$}

\begin{figure}[ht]
 \vspace*{0.cm}
\includegraphics*[width=0.36\textwidth]{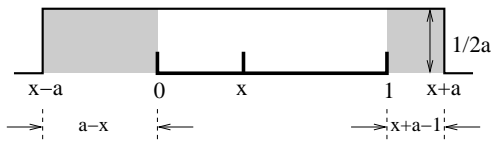}\vskip 0.0in 
\caption{Probability distribution of the uniform random walk after a single
  step.  The shaded regions are the portion of the probability distribution
  outside the unit interval.\label{a1}}
\end{figure}

When $a>1$, the support of the probability distribution necessarily extends
beyond the unit interval after one step.  The residue that remains in the
interval is also uniformly distributed.  These facts allow us to obtain
$R(x)$ and $t(x)$ by simple probabilistic reasoning.  After one step, the
walk jumps past $x=1$ with probability $\frac{x+a-1}{2a}$, and jumps to the
left of the origin with probability $\frac{a-x}{2a}$ (Fig.~\ref{a1}).
Because the remaining probability of $1/2a$ is uniformly distributed within
[0,1], there is a 50\% chance that this residue will eventually exit via
either end.  Thus the exit probability is
\begin{equation}
\label{r1}
R(x)=\frac{x+a-1}{2a}+\frac{1}{2}\frac{1}{2a}=\frac{2(x+a)-1}{4a}\,.
\end{equation}

As expected, the exit probability approaches 1/2, independent of the starting
point, as the average step length becomes large.  Also notice that
$R(1)=\frac{1}{2}+\frac{1}{4a}$; as $a\to 1$ from above, the probability of
exiting the right boundary when starting at $x=1$ is only 3/4.

The survival probability for the walk to remain within the interval after a
single step is simply $1/2a$, and the survival probability after $n$ steps is
then $S(n)=(1/2a)^n$.  Since the first-passage probability for the walk to
first exit the interval at the $n^{\rm th}$ step is $F(n)=S(n-1)-S(n)$, the
mean exit time is
\begin{eqnarray}
\langle t\rangle  &=& \sum_{n=1}^\infty n[S(n-1)-S(n)]\nonumber \\
&=&\sum_{n=0}^\infty S(n)= \frac{1}{1-1/2a}\,.
\end{eqnarray}
As $a\to\infty$, $\langle t\rangle\to 1$, while as $a\to 1$ from above,
$\langle t\rangle \to 2$.  This same value for the exit time can also be
obtained by solving the backward equation for $t(x)$ itself (see below).

\subsection{$a\in[1/2,1]$}

\begin{figure}[ht]
 \vspace*{0.cm}
\includegraphics*[width=0.275\textwidth]{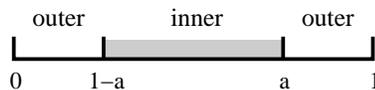}\vskip 0.0in 
\caption{The unit interval for $1/2<a<1$.  A walk starting in $(1-a,a)$ can
  leave the interval via either end in a single step. \label{interval-3}}
\end{figure}

When $1/2<a<1$, the unit interval naturally divides into an inner subinterval
$(1-a,a)$ and outer subintervals $(0,1-a)$, and $(a,1)$
(Fig.~\ref{interval-3}).  If the walk begins in $(1-a,a)$, then the
probability distribution of the walk necessarily extends beyond [0,1] after a
single step, and the exit probability is again given by Eq.~(\ref{r1}).

On the other hand, when the walk starts in the outer subintervals the
recursion formula \eqref{back-r} for the exit probability becomes
\begin{equation}
 R(x) =
\begin{cases}
\frac{1}{2a}\int\limits_0^{x+a} dx' R(x')    &x\in[0,1-a]\,,\\
\frac{1}{2a}\int\limits_{x-a}^{1} dx' R(x') + \frac{x+a-1}{2a}  &x\in[a,1]\,.
\end{cases}
\end{equation}
Differentiating these equations gives
\begin{subequations}
\begin{align}
 2aR'(x) &= R(x+a)    &x\in(0,1-a)\,,\label{h1adif}\\
 2aR'(x) &= 1-R(x-a) &x\in(a,1)\,.\label{h1bdif}
\end{align}
\end{subequations}
When $x\in(0,1-a)$, the second derivative is
\begin{equation}
\label{rmid}
  4a^2R''(x)=2aR'(x+a)=1-R(x),
\end{equation}
where we use the fact that if $x\in(0,1-a)$, then $x+a\in(a,1)$.  The
solution to \eqref{rmid} is
\begin{equation}
\label{h1sola}
 R(x) = 1 + c_1\sin\Big[\frac{1}{2a}(x-c_2)\Big]\,.
\end{equation}

To determine the constants $c_1$ and $c_2$, we first substitute
\eqref{h1sola} into \eqref{h1adif} and also use the fact that $R(x+a)=1-R(x)$
for $x\in(0,1-a)$ to obtain $c_2=(1-a)/2+\pi a/2$.  Second, we match
\eqref{h1sola} and \eqref{r1} at $x=1-a$ to find $c_1$.  The exit probability
for $x\in (0,1-a)$ therefore is
\begin{equation}
\label{r-half}
 R(x) = 1 + \dfrac{\sin\left[\dfrac{1}{2a}\left(x-\dfrac{1-a}{2}\right)
-\dfrac{\pi}{4}\right]}{\left(1-\dfrac{1}{4a}\right)
\sin\left(\dfrac{a-1}{4a}+\dfrac{\pi}{4}\right)}\,.
\end{equation}
The sinusoidal segment of $R(x)$ is visually close to a linear function
(Fig.~\ref{exit}), and the difference between these two functional forms
becomes more clearly visible upon plotting $R'(x)$ versus $x$
(Fig.~\ref{rprime}).

We now compute the mean exit time.  Again, there are two cases to consider:
either the walk begins within $(a,1-a)$ or it begins in the complementary
outer subintervals.  Let us denote by $t_{\rm in}(x)$ and $t_{\rm out}(x)$ as
the mean exit times when the walk starts at a point $x$ in the inner and in
the outer subintervals, respectively.  Then the backward equation
\eqref{back-t} for $t_{\rm in}$ becomes
\begin{eqnarray}
\label{t-in}
t_{\rm in}(x)&\!\!=\!\!&1 \!+\! \frac{1}{2a}\int\limits_0^1\! t(x')\, dx'\nonumber \\
&\!\!=\!\!&1\!+\!\frac{1}{a}\int\limits_0^{1-a} \!t_{\rm out}(x')\, dx'
+\frac{1}{2a}\int\limits_{1-a}^a \!\!t_{\rm in}(x')\, dx'.
\end{eqnarray}
For the last line, we break up the integral into a contribution from the
outer subinterval, with two equal contribution from $(0,1-a)$ and $(a,1)$,
and the inner subinterval $(1-a,a)$.  Notice also from the first line that
$t_{\rm in}(x)$ is independent of $x$.  Thus we define $t_{\rm in}(x)
=\mathcal{T}$, with $\mathcal{T}$ dependent only on $a$.

Similarly, the backward equation for $t_{\rm out}$ is
\begin{equation}
\label{t-out}
t_{\rm out}(x)=1+\frac{1}{2a}\int\limits_0^{x+a} t(x')\, dx'\,.
\end{equation}
Differentiating gives $t_{\rm out}'(x)=t_{\rm out}(x+a)/2a$.  Notice that if
$x\in(0,1-a)$, then $x+a$ is necessarily in $(a,1)$.  Correspondingly, the
backward equation for $t_{\rm out}(x+a)$ gives $t_{\rm out}'(x+a)=-t_{\rm
  out}(x)/2a$.  Thus $t_{\rm out}''(x)=-t_{\rm out}(x)/4a^2$, with solution
\begin{equation}
\label{t-half-out}
t_{\rm out}(x) = \tau_1\cos\left(\frac{x}{2a}\right)+
\tau_2\sin\left(\frac{x}{2a}\right)\,.
\end{equation}
To complete the solution, we need To determine the three unknown constants
$\mathcal{T}$, $\tau_1$, and $\tau_2$.  The solution is straightforward and
the details are given in Appendix~\ref{t-soln}.  From this solution, quoted
in Eq.~\eqref{constants}, we obtain the mean exit times plotted in
Fig.~\ref{exit-time}.  As $a$ decreases, $t(x)$ quickly approaches the
parabolic form of the diffusive limit, but $t(0)$ and $t(1)=t(0)$ remain
strictly greater than zero when $a$ is non zero.

\subsection{$a\in[1/3,1/2]$}

For any $a<1/2$, the exit probability now obeys the generic recursion
formulae:
\begin{equation}
\label{r}
 R(x) =
\begin{cases} \frac{1}{2a}\int\limits_0^{x+a} dx' R(x')  & x\in[0,a]\\ 
\frac{1}{2a}\int\limits_{x-a}^{x+a} dx' R(x')   & x\in[a,1-a]\\
\frac{1}{2a}\int\limits_{x-a}^{1} dx' R(x') + \frac{x+a-1}{2a}  & x\in[1-a,1].
\end{cases}
\end{equation}
For example, the middle equation states that the exit probability starting at
$x$ equals the exit probability after making one step to $x'$---which is
uniformly distributed in the range $-[a,a]$ about $x$---times the exit
probability from $x'$.  The first and third equations account for the
modified range of the single-step distribution if the walk leaves the
interval.

\begin{figure}[ht]
 \vspace*{0.cm}
\includegraphics*[width=0.275\textwidth]{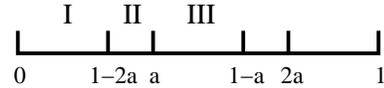}\vskip 0.0in 
\caption{The unit interval with the subregions used to determine the exit
  probabilities for the case $a\in(1/3,1/2)$.\label{interval}}
\end{figure}

We differentiate these integral equations to obtain the more compact form
\begin{subequations}
\begin{align}
 2aR'(x) &= R(x+a)    &x\in(0,a)\,;\label{lesshadif}\\
 2aR'(x) &= R(x+a)-R(x-a)    &x\in(a,1-a)\,;\label{lesshbdif}\\
 2aR'(x) &= 1-R(x-a) &x\in(1-a,1)\,.\label{lesshcdif}
\end{align}
\end{subequations}
To solve these equations for the cases where $a\in[1/3,1/2]$, we should
consider the five subintervals $(0,1-2a)$, $(1-2a,a)$, $(a,1-a)$, $(1-a,2a)$,
and $(2a,1)$ (Fig.~\ref{interval}).  By symmetry, we only need to study the
range $x<1/2$, and we now examine, in turn, subintervals III, I, and II.

{\tt Subinterval III}: For $x\in(a,1-a)$, Eq.~(\ref{lesshbdif}) connects
$R'(x)$ with $R(x+a)$ and $R(x-a)$.  In turn, the equations for $R'(x+a)$ and
$R'(x-a)$ involve $R(x)$ with $x$ again within $(a,1-a)$.  Thus,
\begin{equation}
  4a^2R''(x)=2aR'(x+a)-2aR'(x-a)=1-2R(x)\,,
\end{equation}
with solution
\begin{equation}
\label{thsolc}
 R(x) = \frac{1}{2} + c_1\sin\left[\frac{1}{\sqrt{2}a}\Big(x-\frac{1}{2}\Big)\right]\quad x\in(a,1-a)\,.
\end{equation}
This form automatically satisfies the symmetry condition $R(1/2)=1/2$.  

{\tt Subinterval I}: We obtain the exit probability for $x\in(0,1-2a)$ by
integrating \eqref{lesshadif} and also using the fact that the argument $x+a$
in $R(x+a)$ lies within $(a,1-a)$.  Thus we use the result of
Eq.~\eqref{thsolc} to give
\begin{equation}
\label{thsola}
 R(x) \!=\! \frac{x}{4a} - \frac{c_1}{\sqrt{2}}
\cos\!\left[\frac{1}{\sqrt{2}a}\Big(x\!+\!a\!-\!\frac{1}{2}\Big)\right] + c_2
\quad x\in(0,1-2a)\,.
\end{equation}
To determine the constants $c_1$ and $c_2$, we use the general antisymmetry
condition, $R(y)=1-R(1-y)$, to write \eqref{lesshbdif} in the form
\begin{equation}
\label{rin}
 2aR'(x) = 1-R(1-x-a)-R(x-a)     \quad x\in(a,1-a)\,.
\end{equation}
Now we substitute the solutions \eqref{thsolc} and \eqref{thsola} into
Eq.~(\ref{rin}) and find $c_2=\frac{3}{4}-\frac{1}{8a}$.  Thus
\begin{equation}
\label{thsola2}
 R(x) = \frac{x-1/2}{4a} - \frac{c_1}{\sqrt{2}}
\cos\left[\frac{1}{\sqrt{2}a}\Big(x+a-\frac{1}{2}\Big)\right] + \frac{3}{4}\, .
\end{equation}

{\tt Subinterval II}: Finally, for $x\in (1-2a,a)$, Eqs.~\eqref{lesshadif}
and \eqref{lesshcdif} show that the subintervals $(1-2a,a)$ and $(1-a,2a)$
are coupled only to each other.  Using the antisymmetry of the exit
probability about $x=1/2$, we have
\begin{equation}
 4a^2R''(x) = 1-R(x)    \quad x\in(1-2a,a) \,,
\end{equation}
with solution $R(x)=1+c_3\sin((x-c_4)/2a)$.  We determine $c_4$ from the
condition $R(x) = 1-R(1-x)$ to then give
\begin{equation}
\label{thsolb}
 R(x) = 1 + c_3\sin\left[\frac{1}{2a}\Big(x-\frac{1-a}{2}\Big)-\frac{\pi}{4}\right]\,.
\end{equation}
To obtain the remaining two constants $c_1$ and $c_3$, we 
match Eqs.~\eqref{thsola2} and \eqref{thsolb} at $1-2a$, and \eqref{thsolb} and
\eqref{thsolc} at $a$. These lead to
\begin{subequations}
\begin{align}
c_1 &= \frac{\frac{1}{8a}-\frac{3}{4}+\frac{1}{2}\tan\alpha}{\frac{1}{\sqrt{2}}\cos\beta-\sin\beta\tan\alpha}\label{thc1}\\
c_3 &= \frac{1}{2\cos\alpha} + \frac{\frac{1}{8a}-\frac{3}{4}+\frac{1}{2}\tan\alpha}{\frac{1}{\sqrt{2}}\cos\alpha\cot\beta-\sin\alpha} ,
\end{align}
\end{subequations}
with
\begin{equation}
 \alpha = \frac{3a-1}{4a} + \frac{\pi}{4} , ~~~~~ \beta = \frac{1-2a}{2\sqrt{2}a} .
\end{equation}
For the special case of $a=1/3$, subinterval II disappears so that the
solution consists of \eqref{thsola} and \eqref{thsolc} only, and the relevant
constant in \eqref{thc1} simplifies to
\begin{equation}
 c_1 = \left(4\sqrt{2}\cos\dfrac{1}{\sqrt{8}}-8\sin\dfrac{1}{\sqrt{8}}\right)^{-1}
\end{equation}

\subsection{$a<1/3$}

It is straightforward to treat smaller values of $a$, but the bookkeeping of
the various subintervals becomes increasingly tedious.  However, it is still
possible to infer general properties of the exit probability.  From
Eqs.~\eqref{lesshadif} \& \eqref{lesshbdif}, we have $2aR'(x=a^+)=R(2a)-R(0)$
while $2aR'(x=a^-)=R(2a)$.  Thus $R'(x)$ has a jump of magnitude $R(0)/2a$
when $x$ passes through $a$, as illustrated in Fig.~\ref{rprime} $a\leq 0.4$.
Similarly, consider $R''(x)$ near $x=2a$.  By \eqref{lesshbdif}, $R''(x)$ is
coupled to $R'(x+a)$ and $R'(x-a)$, and the latter derivative has a jump when
its argument equals $a$.  Thus $R''(x)$ has a jump of magnitude $R(0)/4a^2$
when $x$ passes through $2a$.  This pattern continues so that for $a\in
(\frac{1}{n+1},\frac{1}{n})$, the $n^{\rm th}$ derivative of $R(x)$ has a
jump discontinuity as $x$ passes through $na$, while all lower derivatives
are continuous.  Thus $R(x)$ becomes progressively smoother and more linear
in visual appearance for $x$ deeper in the interior of the interval.

\begin{figure}[ht]
 \vspace*{0.cm}
\includegraphics*[width=0.35\textwidth]{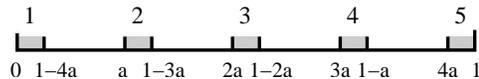}\vskip 0.0in 
\caption{The unit interval (to scale) for the case $\frac{1}{5}<a<\frac{1}{4}$ 
  and the two associated classes of subintervals.  The shaded subintervals
  are labeled sequentially as defined in the text.\label{subintervals}}
\end{figure}

The analytical solution for $R(x)$ can, in principle, be obtained from the
backward equations \eqref{lesshadif}--\eqref{lesshcdif}, for the exit
probability.  When $a<1/2$, these equations naturally partition the unit
interval into two classes of subintervals as shown in
Fig.~\ref{subintervals}.  Nearest-neighbor shaded subintervals are coupled
only to each other by these backward equations, and similarly for the
complementary subintervals.  It is convenient to define $r_k(x)$ as the exit
probability for a walk that begins at $x+(k-1)a$ in the $k^{\rm th}$ shaded
subinterval; that is $r_k(x)=R(x+(k-1)a)$ for $x\in ((k-1)a,ka)$.  With these
conventions, the backward equations for the exit probability for a starting
point in one of the shaded or in one of the unshaded subintervals have the
matrix form:
\begin{equation}
2a \mathbf{r}'(x)=
\left( \begin{array}{rrrrrr}
0 & 1 & 0 &\cdots &&\\
-1 & 0 & 1 &0  &\cdots&\\
0 & -1 & 0& 1&0& \\
\vdots&\vdots&\ddots&\ddots&\ddots&\\
& & 0 & -1 & 0 & 1\\
& && 0& -1& 0\end{array} \right)
\mathbf{r}(x)+ \mathbf{I}\,,
\end{equation}
where $\mathbf{r}(x)$ is the column vector with components
$[r_1(x),r_2(x)\ldots r_n(x)] \!=\! [R(x),R(x\!+\!a)\ldots
R(x\!+\!(n\!-\!1)a)]$ and $\mathbf{I}$ is the column vector with the $n$
components $(0,0,\ldots,0,1)$.

To solve these equations, we note that the eigenvalues of the above matrix
are given by $\lambda_j=2i\cos(\pi j/(n+1))$, with $j=1,2,\ldots,n$
\cite{matrix}.  Thus $r_n(x)$ is, in general, a linear superposition of the
eigenvectors of the matrix; these are sinusoidal functions with arguments
$\lambda_j/2a$.  The actual form of $R(x)$ is then obtained by fixing the
various constants in this eigenvector expansion through matching the
components $r_k$ at appropriate boundary points.

\section{Exit Probability for $a\to 0$}
\label{small-a}

In the limit $a\to 0$, the diffusion approximation becomes increasingly
accurate so that $R(x)$ is very nearly equal to $x$, except in a small region
of the order of $a$ near each boundary (Fig.~\ref{exit}).  This deviation is
more clearly evident when plotting $R'(x)$ versus $x$ (Fig.~\ref{rprime}).
As $a$ gets small, this plot also suggests that a good approximation to
$R'(x)$ will be obtained by solving the exact equation for $R'(x)$ in the
boundary region and treating $R'(x)$ as a constant in the interior of the
interval.

At a zero-order level of approximation, we assume that $R'(x)$ is position
dependent for $x\in(0,a)$ and $x\in(1-a,1)$ and is constant otherwise.  Then
within $(0,a)$, the backward equation $2aR'(x)=R(x+a)$ means that $R'(x)$ is
a linear function (and similarly for $x\in(1-a,1)$).  Thus we make the ansatz
\begin{equation}
\label{r-asymp}
R(x)\approx
\begin{cases}\frac {1}{2}(1-s)+ s x  & x\in(a,1-a) \\
r_0+r_1 x+r_2 x^2 & {\rm otherwise}\,,
\end{cases}
\end{equation}
with $s$ and the $r_i$ to be determined.  We expect the slope $s$ of $R(x)$
in the interior of the interval to approach 1 as $a\to 0$ to recover $R(x)\to
x$ in this limit.  The form of the constants in the first line also ensure
the obvious special case $R(1/2)=1/2$.  Similarly, the linear form for
$R'(x)$ in the boundary regions roughly corresponds to what is seen in
Fig.~\ref{rprime}.

\begin{figure}[ht]
 \vspace*{0.cm} \includegraphics*[width=0.4\textwidth]{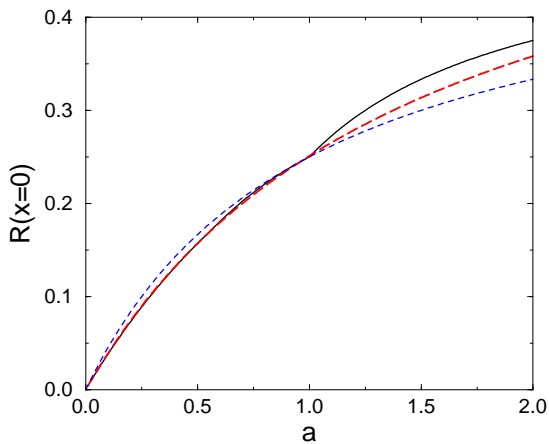}\vskip 0.1in
 \caption{Exit probability $R(x=0)$ versus $a$, based on $5\times 10^8$
   walks for 200 equally-spaced values of $a$ between 0 and 2. Shown dotted
   is the result of our zeroth-order asymptotic approximation
   $R(0)=a/[2(1+a)]$, while the dashed curve is the result of the first-order
   approximation.
   \label{r0}}
\end{figure}
 
We determine the 4 unknowns in the above asymptotic approximation for $R(x)$
by the following conditions: (i) $2aR'(x)=R(x+a)$ must be satisfied in the
region $x\in (0,a)$ (this gives two conditions---one for the linear term and
one for the constant term), (ii) $R(x)$ is continuous, (iii) the
discontinuity in $R'(x)$ at $x=a$ equals $R(0)/2a$, as follows from
Eqs.~\eqref{lesshadif} and \eqref{lesshbdif}.  Applying these conditions
gives, after some simple calculation,
\begin{eqnarray*}
s\!=\!\frac{1}{1+a};~
r_0\!=\!\frac{1}{2}\frac{a}{1+a};~
r_1\!=\!\frac{3}{4}\frac{1}{1+a};~
r_2\!=\!\frac{1}{4a}\frac{1}{1+a}\,.
\end{eqnarray*}

A much better approximation is obtained by treating $R(x)$ exactly in the
domains $(0,a)$ and $(a,2a)$, and then assuming that $R'(x)$ is constant
otherwise.  Thus for $x>2a$, $R(x)$ is given by the first line of
Eq.~\eqref{r-asymp}, while within $(0,a)$ and $(a,2a)$, the governing
backward equations for $R(x)$ are
\begin{eqnarray}
2aR'(x)=
\begin{cases}
R(x+a)  & x<a \\ R(x+a)-R(x-a) & x>a
\end{cases}
\end{eqnarray}
For $x<a$, we iterate the first equation to give $4a^2R''(x)= R(x+2a)-R(x)$
and make use of the assumption that $R'(x)=s$ for $x>2a$.  This leads to
the approximation
\begin{equation}
\label{a0-AB}
R(x)= d_1\cos\left(\frac{x}{2a}\right) + d_2\sin\left(\frac{x}{2a}\right)+
s(x\!+\!2a)+\frac{1}{2}(1\!-\!s)\,.
\end{equation}
Similarly, to obtain $R(x)$ in the region $(a,2a)$, we integrate the backward
equation $2aR'(x)=R(x+a)-R(x-a)$ and again use the fact that the argument
$x+a$ in $R(x+a)$ is beyond $2a$, so that $R(x+a)$ is a linear function.
This integration leads to 
\begin{equation}
\label{a0-C}
R(x)= -d_1\sin\left(\frac{x-a}{2a}\right)+ d_2\cos\left(\frac{x-a}{2a}\right)+d_3\,.
\end{equation}

We determine the 4 coefficients in these two forms for $R(x)$ by requiring
that at $x=a$, $R$ is continuous and the first derivative has a jump of
magnitude $R(0)/2a$, while at $x=2a$, both $R$ and $R'$ are continuous.  The
resulting formulae are given in Appendix~\ref{a0-detail}.  Fig.~\ref{rprime}
shows the result of this small-$a$ approximation for $R'(x)$.  The agreement
between this asymptotic approximation and the numerical results is extremely
good.

As a further test of the accuracy of this approach, we show the
numerically-obtained dependence of $R(0)$ on $a$ together with our zeroth-
and first-order approximations for $R(0)$ (Fig.~\ref{r0}).  As already
mentioned in Sec.~\ref{general}, $R(0)$ is greater than zero for any $a>0$
because there is a non-zero chance that a walk exactly at the left boundary
can still exit via the right boundary.  In the limit of small $a$, our
asymptotic approach for $R(0)$ closely approximates the data.

\section{Concluding Remarks}

An enigmatic feature of the uniform random walk (URW) is that its
first-passage properties in a finite interval are not described in terms of a
radiation boundary condition \cite{fpp,W}.  This boundary condition is
$c'=c/\kappa$, where $\kappa$ is the extrapolation length.  This condition
can be interpreted as partial absorption and partial reflection when a walk
hits the boundary.  In the case of a URW that starts at one absorbing
boundary, there is a non-zero chance for the walk to exit via the opposite
boundary.  This incomplete absorption should be equivalent to partial
reflection from the initial boundary, which should be described by a
radiation boundary condition.  Indeed, the probability distribution of a URW
in a semi-infinite interval $x>0$ with absorption at $x=0$ closely
approximates that obtained for pure diffusion with a radiation boundary
condition.  However, in the finite interval the radiation boundary condition
gives $R(x)$ as the linear function $R(x)=\frac{x+\kappa}{1+2\kappa}$ which
does not account for the anomalous behavior observed near the edges of the
interval.

What we do find is that the first-passage properties of the URW in a finite
interval exhibit curious commensuration effects as $a$ passes through
$1,\frac{1}{2},\frac{1}{3}\,\ldots$.  For $a\in (\frac{1}{n+1},\frac{1}{n})$,
the $n^{\rm th}$ derivative of $R(x)$ has a jump discontinuity as $x$ passes
through $na$, while all lower derivatives are continuous.  The exit time has
corresponding singular behaviors.  For small $n$, we have computed the exit
probability and the mean exit time exactly by a direct probabilistic
approach.  In the limit $a\to 0$, the diffusion approximation becomes
increasingly accurate, except when the starting point is close to either
boundary where the exit probability continues to exhibit non-diffusive
effects.  In the limit of small $a$, we constructed an approximation of
treating the exit probability exactly within the boundary region, an approach
that gives extremely accurate results.

\acknowledgments{We thank G. Huber for stimulating our interest in this
  problem, as well as H. J. Hilhorst and G. H. Weiss for literature advice.
  We also acknowledge financial support from the Swiss National Science
  Foundation under fellowship 8220-067591NSF (TA) as well as US National
  Science Foundation grants DMR0227670 and DMR0535503 (SR).  }

\appendix

\section{Mean Exit Time for $a>1/2$}
\label{t-soln}

To complete the solution for the mean exit time, we substitute $t_{\rm in}+a$
and $t_{\rm out}$ quoted in Eq.~\eqref{t-half-out} into Eqs.~\eqref{t-in} and
\eqref{t-out}.  The former equation becomes
\begin{eqnarray}
\mathcal{T}=1+\frac{1}{a}\int_0^{1-a} t_{\rm out}(x')\, dx' + \frac{2a-1}{2a}\, \mathcal{T}\,.
\end{eqnarray}
This can be rewritten as
\begin{eqnarray}
\label{AA}
\mathcal{T}=2a +2T\,,
\end{eqnarray}
where 
\begin{eqnarray}
T&=&\int_0^{1-a} t_{\rm out}(x')\, dx'\nonumber \\
&=& \int_0^{1-a} \tau_1\cos\left(\frac{x'}{2a}\right)
+ \tau_2\sin\left(\frac{x'}{2a}\right)\,dx'\nonumber \\
&=& 2a\tau_1\sin\left(\frac{1-a}{2a}\right)
-  2a\tau_2\cos\left(\frac{1-a}{2a}\right)\,.
\end{eqnarray}

For Eq.~\eqref{t-out}, we evaluate it at $x=0$.  This gives
\begin{eqnarray*}
t_{\rm out}(0)=1+\frac{1}{2a}\int_0^a t(x')\, dx'\,.
\end{eqnarray*}
Then using  Eq.~\eqref{t-half-out} for $t_{\rm out}(0)$ and separating the
integral into the inner and outer subintervals, we have
\begin{eqnarray}
\label{BB}
\tau_1=1+\frac{1}{2a}\,T + \left(1-\frac{1}{2a}\right) \mathcal{T}\,.
\end{eqnarray}

Finally, we equate $t_{\rm in}$ and $t_{\rm out}$ at $x=1-a$, where the inner
and outer subintervals meet.  This gives
\begin{eqnarray}
\label{CC}
\mathcal{T}= \tau_1\cos\left(\frac{1-a}{2a}\right)
+  \tau_2\sin\left(\frac{1-a}{2a}\right)\,.
\end{eqnarray}
The conditions \eqref{AA}, \eqref{BB}, and \eqref{CC} provide the three
independent equations
\begin{eqnarray*}
\mathcal{T}&\!=\!& 2a  +4a\tau_1\sin\left(\frac{1\!-\!a}{2a}\right)
- 4a\tau_2\left[\cos\left(\frac{1\!-\!a}{2a}\right)\!-\!1\right]\nonumber \\
\tau_1&\!=\!&1+a\tau_1\sin\left(\!\frac{1\!-\!a}{2a}\!\right)
-\tau_2\left[\cos\left(\frac{1\!-\!a}{2a}\right)\!-\!1\right] \!+\!
\left(\!1\!-\!\frac{1}{2a}\!\right) \\
\mathcal{T}&\!=\!& \tau_1\cos\left(\frac{1\!-\!a}{2a}\right)
+\tau_2\cos\left(\frac{1\!-\!a}{2a}\right)\,.
\end{eqnarray*}
for the unknown coefficients $\mathcal{T}$, $\tau_1$, and $\tau_2$.

To express the solution succinctly, let $z\equiv \frac{1-a}{2a}$ and
$\epsilon\equiv \left(1-\frac{1}{2a}\right)$.  Further, define
\begin{eqnarray*}
\alpha&=&4a\sin z-\cos z\quad \quad \beta=4a(\cos z-1)+\sin z\\
\gamma&=&\sin z -1 +\epsilon \cos z \quad \delta=\cos z -1 -\epsilon \sin z\,.
\end{eqnarray*}
Then the constants are
\begin{eqnarray}
\label{constants}
\tau_1\!=\!\frac{2a\delta\!-\!\beta}{\gamma\beta\!-\!\alpha \delta}\quad 
\tau_2\!=\!\frac{2a\!+\!\alpha\tau_1}{\beta}\quad
\mathcal{T}\!=\!\tau_1\cos z\!+\! \tau_2\sin z.
\end{eqnarray}
The result of this solution is shown in Fig.~\ref{exit-time}.

\section{Coefficients of the Exit Probability for $a\to 0$}
\label{a0-detail}

At $x=a$, $R(x)$ is continuous, while $R'(x)$ has a discontinuity of size
$R(0)$.  Similarly, at $x=2a$, both $R(x)$ and $R'(x)$ are continuous.  From
Eqs.~\eqref{a0-AB} and \eqref{a0-C}, we thus have the conditions:
\begin{eqnarray}
d_1\cos\frac{1}{2} +d_2\sin\frac{1}{2} +3s
a+\frac{1}{2}(1-s)=d_2+d_3\nonumber \\
-d_1\sin\frac{1}{2} +d_2\cos\frac{1}{2} =\frac{1}{2}(1-s)\nonumber \\
-d_1\sin\frac{1}{2} +d_2\cos\frac{1}{2}+d_3= +2s a+\frac{1}{2}(1-s)\nonumber\\
-d_1\cos\frac{1}{2} -d_2\sin\frac{1}{2}=2s a\,.
\end{eqnarray}
The solution to these equations are:
\begin{eqnarray}
d_1&=& \frac{(1-u)/2}{(u-1)(\frac{1}{4a}\cos\frac{1}{2}+\sin\frac{1}{2})
+(\cos\frac{1}{2}-\frac{1}{4a}\sin\frac{1}{2})v} \nonumber\\
s&=& -\frac{d_1}{2a}\cos\frac{1}{2}-\frac{d_2}{2a}\sin\frac{1}{2}\nonumber\\
d_2&=&-\frac{vd_1}{u-1}\nonumber\\
d_3&=&2s a +\frac{1}{2}(1-s)+d_1\sin\frac{1}{2} -d_2\cos\frac{1}{2}\,,
\end{eqnarray}
with 
\begin{eqnarray}
u=\cos\frac{1}{2}+\frac{1}{2}\sin\frac{1}{2}\quad
v=\frac{1}{2}\cos\frac{1}{2}-\sin\frac{1}{2}\,.
\end{eqnarray}


\begin{thebibliography}{99}
  
\bibitem{general} G. Adam and M. Delbruck, in {\it Structural Chemistry and
    Molecular Biology}, eds.\ A. Rich and N. Davidson (Freeman, San
  Francisco, CA, 1968); O. G. Berg, R. B. Winter, and P. H. Von Hillel,
  Biochemistry {\bf 20}, 6929 (1981).
  
\bibitem{C} M. Coppey, O. Benichou, R. Voituriez, and M. Moreau,
Biophys.\ J. {\bf 87}, 1640 (2004). 

\bibitem{M} R. Murugan, Phys.\ Rev.\ E {\bf 69}, 011911 (2004).
  
\bibitem{MM} M. Montero, J. Perell\'o, J. Masoliver, F. Lillo, S. Miccich\`e,
  and R. N. Mantegna, Phys.\ Rev.\ E {\bf 72}, 056101 (2005).

\bibitem{DEH} B. Dybiec, E. Gudowska-Nowak, and P. H\"anggi,
  cond-mat/0512492.

\bibitem{MCZ} S. N. Majumdar, A. Comtet, and R. M. Ziff, J. Stat.\ Phys. (in
  press).

\bibitem{LS} K. Kakatos-Lindenberg and K. E. Shuler, J. Math.\ Phys. {\bf
    12}, 633 (1971).

\bibitem{L} K. Lindenberg, J. Stat.\ Phys.\ {\bf 10}, 485 (1974). 

\bibitem{WS} G. H. Weiss and A. Szabo, Physica {\bf 119A}, 569 (1983).

\bibitem{BPBW} M. Bogu\~n\'a, S. Pajevic, P. J. Basser, and G. W. Weiss,
  New J. Phys.\ {\bf 7}, 24 (2005).


\bibitem{feller} W. Feller {\it An Introduction to Probability Theory and Its
    Applications}, (Wiley, New York, 1968).

\bibitem{fpp} S. Redner, {\it A Guide to First-Passage Processes}, Cambridge
  University Press, New York (2001); N. G. van Kampen, {\it Stochastic
    Processes in Physics and Chemistry}, 2nd ed.  (North-Holland, Amsterdam,
  1997).

\bibitem{Arf} See {\it e.g.}, G. B. Arfken and H. J. Weber, {\it Mathematical
    Methods for Physicists}, $4^{\rm th}$ ed.\ (Academic Press, San Diego,
  1995).

\bibitem{martin} See P. G. Doyle and J. L. Snell, {\it Random Walks and
    Electric Networks}, (Carus Mathematical Monographs, no.~22, Mathematical
  Association of America, 1984) [also reposted as math.PR/0001057] for a
  physical description of martingales.
  
\bibitem{matrix} R. Bellman, {\it Introduction to Matrix Analysis}
  (McGraw-Hill, New York, 1970); see also K. Kang and S. Redner, J. Chem.\
  Phys.\ {\bf 80}, 2752 (1984).
  
\bibitem{W} G. H. Weiss, {\it Aspects and Application of the Random Walk}
  (North-Holland, Amsterdam, 1994).

\end{thebibliography}
\end{document}